\documentclass[aps,prl,twocolumn,superscriptaddress,showpacs]{revtex4}

\usepackage{graphics}  
\usepackage{bm}        
\usepackage{amssymb}   
\usepackage{color} 

\newcommand{\strike}[1]{\textcolor{red}{*}} 

\begin{document}

\title{Observing the decay of orbital angular momentum entanglement, \\ through experimentally simulated turbulence}

\author{Alpha Hamadou Ibrahim}
\affiliation{CSIR National Laser Centre, PO Box 395, Pretoria 0001, South Africa}
\affiliation{University of Kwazulu-Natal, Private Bag X54001, 4000 Durban, South Africa}
\author{Filippus S. Roux} 
\affiliation{CSIR National Laser Centre, PO Box 395, Pretoria 0001, South Africa}
\author{Sandeep Goyal} 
\affiliation{University of Kwazulu-Natal, Private Bag X54001, 4000 Durban, South Africa}
\author{Melanie McLaren} 
\affiliation{CSIR National Laser Centre, PO Box 395, Pretoria 0001, South Africa}
\affiliation{Laser Research Institute, University of Stellenbosch, Stellenbosch 7602, South Africa}
\author{Thomas Konrad} 
\affiliation{University of Kwazulu-Natal, Private Bag X54001, 4000 Durban, South Africa}
\affiliation{National Institute of Theoretical Physics, Durban Node, South Africa}
\author{Andrew Forbes} 
\affiliation{CSIR National Laser Centre, PO Box 395, Pretoria 0001, South Africa}
\affiliation{University of Kwazulu-Natal, Private Bag X54001, 4000 Durban, South Africa}
\affiliation{Laser Research Institute, University of Stellenbosch, Stellenbosch 7602, South Africa}

\begin{abstract}
We study the evolution of an orbital angular momentum (OAM) entangled bipartite photonic state for the case where one of the photons propagates through Kolmogorov turbulence, using the concurrence as a measure of entanglement. Quantum state tomography was performed to reconstruct the two qubit density matrices for a range of scintillation strengths. Our results give the first direct experimental confirmation of the existing theories for decay of entanglement due to atmospheric turbulence. We also show how the modal scattering increases with increasing scintillation and we discuss the impact of the scale at which entanglement dissipates due to atmospheric turbulence on free-space quantum communication.
\end{abstract}



\maketitle  

\noindent
Laguerre-Gaussian (LG) modes have recently attracted much attention within the quantum information community, mainly because their infinite-dimensional Hilbert space allows information processing with higher dimensional quantum states (qudits) \cite{MT} for use in higher dimensional quantum key distribution \cite{SG} and long-range quantum communication \cite{P2}, among others. An LG beam with azimuthal index $\ell$ carries an orbital angular momentum (OAM) of $\ell\hbar$ per photon \cite{ABSW, FHRD}.

Quantum entanglement is an important resource for quantum information processing and quantum communication, but suffers decay when encountering a noisy channel \cite{konrad1,tiersch}, such as atmospheric turbulence. Various aspects of OAM modes propagating through turbulence have been considered theoretically, including the detection probability of OAM modes \cite{P1b, P2, TB}, attenuation and crosstalk among multiple OAM channels \cite{JA} and the decay of entanglement for bipartite qubits \cite{P1,FSR}. While most of these studies \cite{P1b, TB, P1} are based on a single phase screen approximation \cite{P2}, which is only valid in weak scintillation, the case of arbitrary scintillation strength, which requires a multiple phase screen approach, has also been considered with the aid of an infinitesimal propagation equation (IPE) \cite{FSR}. These investigations show that OAM entanglement is more robust when the beam waist radius $w_0$ is small compared to the turbulence coherence length, which is given by the Fried parameter \cite{FRI} $r_0=0.185(\lambda^2/C_n^2 z)^{3/5}$, where $C_n^2$ is the refractive index structure constant, $z$ is the propagation distance and $\lambda$ is the wavelength. The studies also indicate that entanglement decays comparatively slower for larger OAM.

From an experimental point of view, the problem of atmospherically induced decay of OAM entanglement has thus far received little attention. It was shown, using coincidence counts, that the number of entangled modes (the Shannon dimensionality) decreases with increasing scintillation \cite{BJP}. In other studies \cite{MehulMalik2012, BRodenburg2012a} the crosstalk among OAM modes has been measured, using a single phase screen to simulate the turbulence experimentally. None of these experiments directly addressed the dissipation of entanglement due to atmospheric scintillation.

In this letter we present the first experimental results on the decay of OAM entanglement between two qubits due to atmospheric turbulence. To allow quantitative comparison between our results and the existing theories that are based on \cite{P2}, we consider only qubits and implement the turbulence as a phase-only distortion on a single thin phase screen, using a spatial light modulator (SLM). We compute the concurrence \cite{W} directly from the density matrices obtained by employing full state tomography \cite{QST} to observe the entanglement dissipation as a function of scintillation strength.

\begin{figure}[ht]
\centerline{\includegraphics{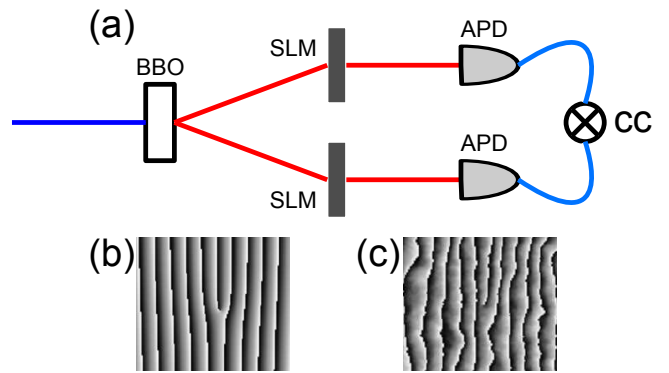}}
\caption{Experimental setup used to detect the OAM eigenstates after SPDC (a). The plane of the crystal was imaged onto two separate SLMs where the modes were selected, as shown in (b) and (c). The SLM planes were then imaged to the inputs of single-mode fibers, connected to APDs. Random phase fluctuations were added to the mode on one of the SLMs to simulate Kolmogorov turbulence (c).}
\label{fig:setup}
\end{figure}

A diagram of our experimental setup is shown in Fig.~\ref{fig:setup}(a). A mode-locked laser source with a wavelength of 355 nm and an average power of 350 mW was used to pump a 0.5 mm-thick type I BBO crystal to produce non-collinear, degenerate photon pairs via spontaneous parametric down-conversion (SPDC). Down-converted photon pairs are naturally entangled in terms of the OAM basis due to the conservation of OAM in SPDC \cite{zeil1}. The plane of the crystal was imaged onto two separate SLMs in the signal and idler beams, respectively. The SLMs served to perform projective measurements for quantum state tomography \cite{QST} by selecting particular pairs of modes for detection [see Fig.~\ref{fig:setup}(b)]. The atmospheric turbulence was simulated by adding random phase fluctuations to the phase function of one of the SLMs [see Fig.~\ref{fig:setup}(c)]. The SLM planes were re-imaged and coupled into single-mode fibers, which extract the pure Gaussian mode from the incident field. Avalanche photo diodes (APDs) that were connected to the fibers registered the photon pairs via a coincidence counter (CC). All measured coincidence counts were accumulated over a 10 second integration time with a gating time of 10 ns.

The random phase fluctuation $\theta(x,y)$ that was added to the one SLM was obtained by the method of filtering white Gaussian noise \cite{MF1, KNEPP1983}:
\begin{equation}
\theta(x, y) = {\cal F}^{-1} \left\{\frac{\chi({\bf k}_\perp) \left[\Phi_\theta({\bf k}_\perp) \right]^{1/2}}{\Delta_{k}} \right\} .
\label{eqn:phScr}
\end{equation}
Here ${\cal F}^{-1}\{\cdot\}$ is the two-dimensional inverse Fourier transform, ${\bf k}_\perp$ is the two-dimensional wave vector in the transverse Fourier domain, $\Delta_{k}$ is the sampling interval in the frequency domain and $\chi({\bf k}_\perp)$ is a frequency domain zero-mean Gaussian pseudo-random complex function, obeying $\chi^*({\bf k}_\perp)=\chi(-{\bf k}_\perp)$, because $\theta(x, y)$ is real-valued. The phase power spectral density is related to the refractive index power spectral density through $\Phi_\theta({\bf k}_\perp) = 2\pi k_0^2 z \Phi_n({\bf k}_\perp,0)$, where $k_0$ is the wavenumber ($2\pi/\lambda$). For comparison with existing theories, we used the Kolmogorov spectrum $\Phi_n^K( k) = 0.033~C_n^2 k^{-11/3}$ \cite{K, AP} and to ensure that the random phase functions can reproduce the Kolmogorov structure function we added subgrid sampling points to the Fourier domain representation of the phase function \cite{RGLane1992}. Our experimentally simulated scintillation strengths ranged from $w_0/r_0 = 0$ to $4$, in 21 increments. For each scintillation strength we performed 26 realizations and a state tomography \cite{QST} was performed for each realization to reconstruct the bipartite qubit density matrix.

The entanglement is quantified by the concurrence \cite{W} ${\cal C}(\rho) = \max\{0,\sqrt{\lambda_1}-\sqrt{\lambda_2}-\sqrt{\lambda_3}-\sqrt{\lambda_4}\}$, where the $\lambda_n$'s are the eigenvalues, in decreasing order, of the matrix $\rho(\sigma_y\otimes\sigma_y)\rho^*(\sigma_y\otimes\sigma_y)$. Here $^*$ represents the complex conjugate and $\sigma_y$ is the Pauli $y$-matrix.

In normalized coordinates the LG modes are
\begin{equation}
M^{\rm LG}_{\ell p} = {\cal N} r^{|\ell|} \exp({\rm i} \ell \phi) L^{|\ell|}_p \left(\frac{2 r^2}{1 + t^2} \right) \exp \left(\frac{-r^2}{1 - {\rm i} t} \right) ,
\label{lgmodus}
\end{equation}
where $L_p^{|\ell|}(\cdot)$ represents the generalized Laguerre polynomials with azimuthal index $\ell$ and radial index $p$, $\phi$ is the azimuthal angle, $r=(x^2+y^2)^{1/2}/w_0$ and $t=z/z_R$, with $z_R$ being the Rayleigh range ($\pi w_0^2/\lambda$). The normalization and Gouy phase factors are contained in
\begin{equation}
{\cal N} = { (1 + {\rm i} t)^p \over (1 - {\rm i} t)^{p + |\ell| +1}} \left[ \frac{p! 2^{|\ell| + 1}}{\pi (p + |\ell|)!} \right]^{1/2} .
\end{equation}

The full spatial entanglement of down-converted photons is experimentally accessible, if all LG modes with arbitrary $\ell$ and $p$ indices are considered \cite{SEL}. However, here only LG modes with $p=0$ are measured. This allows us to express the state of a down-converted photon pair as $|\psi\rangle = \sum_{\ell} c_{\ell} |\ell\rangle |-\ell\rangle$, where $|\ell\rangle$ represents the LG modes with $p=0$ and $|c_{\ell}|^{2}$ denotes the probability amplitude for detecting the signal and the idler photons in the state $|\ell\rangle$ and $|-\ell\rangle$, respectively. Here we only consider qubits defined in terms of the OAM states with $\ell = \pm 1$.

\begin{figure}[ht]
\centerline{\includegraphics{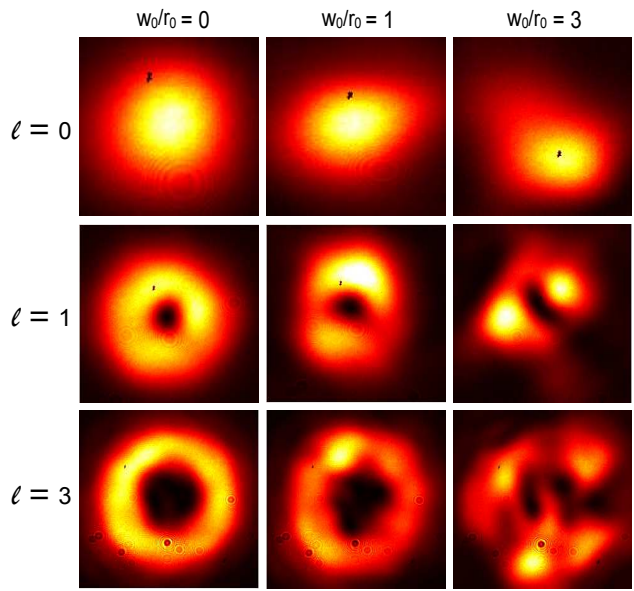}}
\caption{Experimentally measured intensity distributions of LG beams for $\ell = 0,1,3$ and $p = 0$, via back-projection from the fiber onto the SLM (see text). First column without turbulence ($w_0/r_0=0$). Second column for $w_0/r_0 = 1$ and the last column for $w_0/r_0 = 3$. }
\label{fig:BP}
\end{figure}

Any classical LG beam propagating through turbulence is distorted by random phase modulations that, after some propagation, turn into intensity fluctuations. Examples of such intensity distortions are shown in Fig.~\ref{fig:BP}. These images were obtained in the plane of the BBO crystal with back-projected illumination \footnote{The APD is replaced by a laser diode, shining light backward through the fibre to illuminate the SLM.} of the SLM that contains the phase fluctuations in the experimental setup shown in Fig.~\ref{fig:setup}(a). The distortions of the LG mode implies that the energy of the original LG mode is scattered into other LG modes.

A similar scattering process occurs in the case of an entangled quantum state propagating through turbulence. Given an initial pure state $|\ell_0\rangle \langle \ell_0|$, after turbulence it becomes $|\ell_0 \rangle \langle \ell_0| \rightarrow \sum_{mn} \rho_{mn} |\ell_m\rangle \langle \ell_n|$ (ignoring the $p$-index), where $\rho_{mn}$ are the density matrix elements.

\begin{figure}[ht]
\centerline{\includegraphics{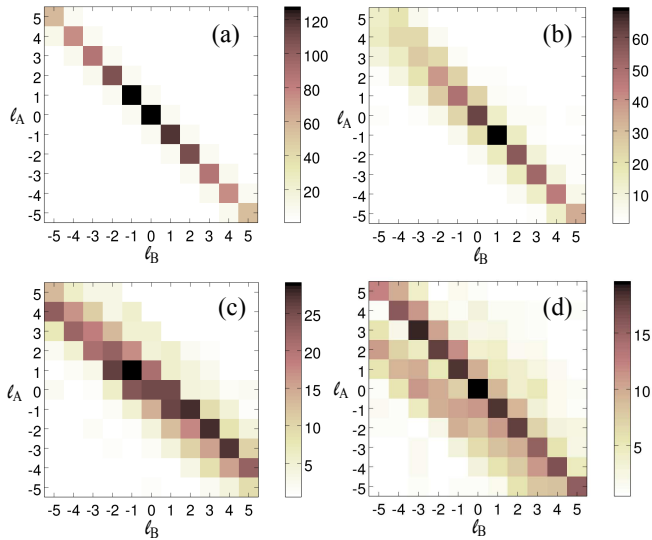}}
\caption{Mode scattering under the effect of turbulence given by the coincidence counts for simultaneous measurements of modes with azimuthal index $\ell_A$ in the signal beam and $\ell_B$ in the idler beam. (a) With no turbulence, only anti-correlated coincidences are observed. As the scintillation strength increases, the mode scattering becomes more pronounced: (b) $w_0/r_0 = 1$, (c) $w_0/r_0 = 2$ and (d) $w_0/r_0 = 3$.}
\label{fig:BW}
\end{figure}

We have experimentally measured this modal scattering for different scintillation strengths shown as modal diagrams in Fig.~\ref{fig:BW}. Without scintillation the modal diagram displays a perfect correlation between $\ell_A$ and $-\ell_B$, as seen from the strong diagonal in Fig.~\ref{fig:BW}(a), which is indicative of the entanglement of the down-converted photon state. As the scintillation is increased, one can see from Figs.~\ref{fig:BW}(b) to (d) that the components of the quantum state are gradually scattered away from the diagonal, first to the neighboring locations and then to locations further away. This indicates that the mode coupling is strongest for neighboring modes and decreases as the mode separation increases.

\begin{figure}[ht]
\centerline{\includegraphics{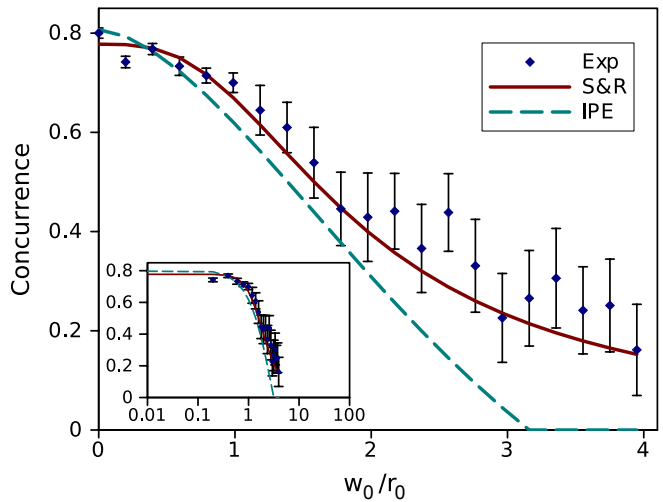}}
\caption{Concurrence as a function of scintillation strength ($w_0/r_0$). The (blue) diamonds represent the average concurrence obtained from the experimental measurements, together with their standard deviations, shown as error bars. The solid (red) line represents the theoretical curve obtained with the single phase screen approximation (weak scintillation) and the dashed (green) line is the theoretical curve obtained from the IPE. The inset shows the same graph on a logarithmic horizontal axis to emphasize the scale of entanglement decay.}
\label{fig:conc}
\end{figure}

Our main result is the experimental observation of the decay of entanglement as a function of the scintillation strength, given by $w_0/r_0$. In Fig.~\ref{fig:conc} we compare the experimentally obtained concurrence values with the theoretical concurrence curves computed within the single phase screen approximation, as done by Smith and Raymer (S\&R) \cite{P1}, and with the IPE \cite{FSR}. Since the experiment imposes scintillation on only one of the two photons, the calculation of the two theoretical curves was also done for only one photon passing through turbulence. The experimental results were obtained from numerous quantum state tomography measurements, from which more than 500 individual density matrices were reconstructed (26 realizations for 21 points). These density matrices were used to compute the concurrence. Care was taken to remove any negative eigenvalues that occur for these density matrices \footnote{We added an identity matrix times the absolute value of the most negative eigenvalue to the density matrix and renormalized the result. If the error bars of the resulting eigenvalues, computed from Poisson statistics, still pushed below zero, we adjusted the mean and standard deviations of these eigenvalues so that they remain above zero.}. Such negative eigenvalues are caused by measurement errors --- fluctuations in the coincidence counts caused by fluctuations in the photon number statistics. Each of the points in Fig.~\ref{fig:conc} represents an average of 26 such concurrence values and the error bars indicate the associated standard deviations. Note that the experimental values indicate a maximum concurrence without turbulence of about ${\cal C}=0.8$. As a result we adjusted the theoretical curves by an overall factor to provide the best fit with the experimental data. The lower concurrence is due to experimental imperfections and photon statistics. 

There is a good agreement between the experimental results and the S\&R curve. This is the first experimental confirmation of these theoretical results as far as we are aware. The experimental results deviate from the IPE curve, especially for larger values of $w_0/r_0$. This is not unexpected, in view of the fact that the single phase screen approximation is only valid in weak scintillation conditions. It should be noted that both the experimental setup and the S\&R curve represent a single phase screen approach, while the IPE does not make this approximation. The decrease in entanglement (Fig.~\ref{fig:conc}) is consistent with the modal scattering observed in Fig.~\ref{fig:BW}.

The most pertinent point of the results shown in Fig.~\ref{fig:conc} is the fact that it provides experimental confirmation of the scale of entanglement decay. Viewing the curves on a logarithmic horizontal scale, as shown in the inset in Fig.~\ref{fig:conc}, one finds that the decay of entanglement all happens within an order of magnitude around the point where $w_0/r_0 = 1$. The fact that both theoretical curves and the experimental data share the same scale indicates that, whether or not one uses a single phase screen approximation does not make any difference to that scale. Substituting the expression for the Fried parameter into $w_0/r_0 = 1$, we derive the distance scale at which atmospherically induced decay of entanglement occurs:
\begin{equation}
L_{\rm dec} = {0.06\lambda^2\over w_0^{5/3} C_n^2} .
\label{zskaal}
\end{equation}
For practical free-space quantum communication systems the distances between repeaters would need to be shorter than $L_{\rm dec}$. Although this distance would often be shorter than the Rayleigh range \cite{BJP}, in weak turbulence this distance could be larger than the Rayleigh range.

Interestingly, the scale where $w_0/r_0 = 1$ roughly coincides with the transition region between weak and strong scintillation. Scintillation is classified as either weak or strong based on the Rytov variance, which is defined as $\sigma_R^2 = 1.23 C_n^2 k_0^{7/6} z^{11/6}$. For plane waves, strong scintillation is said to exist when $\sigma_R^2>1$ \cite{AP} and for Gaussian beams it exists when $\sigma_R^2 > (t+1/t)^{5/6}$ \cite{MRA}.

\begin{figure}[ht]
\centerline{\scalebox{1}{\includegraphics{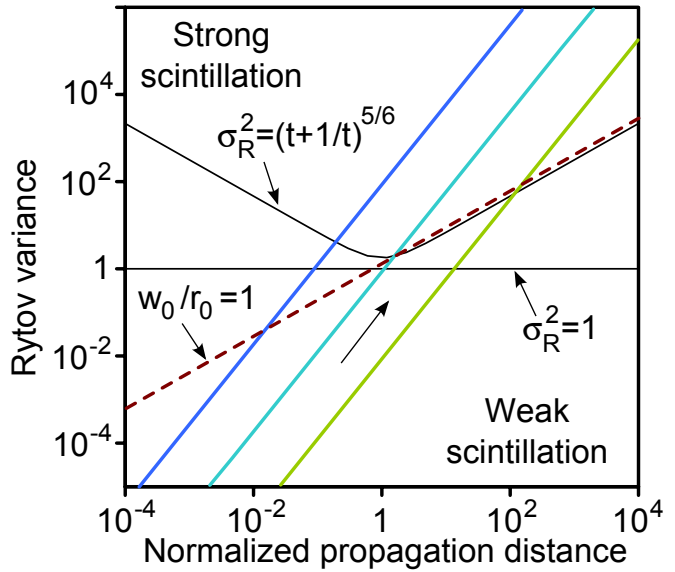}}}
\caption{A diagram showing scintillation strength in terms of $\sigma_R^2$ as a function of $t$ on a log-log scale. Weak scintillation lies toward the bottom and strong scintillation toward to top, with the boundary between these two regions shown for both plane waves ($\sigma_R^2=1$) and Gaussian beams [$\sigma_R^2=(t+1/t)^{5/6}$]. The dashed line ($w_0/r_0=1$) is roughly where the concurrence approaches zero and the three slanted colored lines indicate the increase in $\sigma_R^2$ with propagation distance for different strengths of turbulence (left to right:\ $C_n^2=\{10^{-12}, 10^{-14}, 10^{-16}\}$ m$^{2/3}$).}
\label{rytovfig}
\end{figure}

The diagram in Fig.~\ref{rytovfig} shows the different regions in terms of the Rytov variance as a function of the normalized propagation distance $t$. For any particular optical beam propagating through turbulence, the value of $\sigma_R^2$ is proportional to $t^{11/6}$. Three such lines are shown in Fig.~\ref{rytovfig} for different turbulence strengths. One can see that these lines start off in the region of weak scintillation at the bottom of the diagram and then move up toward the region of strong scintillation as the beams propagate further. Eventually these lines cross the boundary into the region of strong scintillation [$\sigma_R^2=(t+1/t)^{5/6}$]. However, at the same time (for weak turbulence) or even before it (for strong turbulence) it also crosses the line where $w_0/r_0=1$ (dashed line in Fig.~\ref{rytovfig}), which is the scale where entanglement decays. The dashed line is obtained by expressing the Rytov variance in terms of $w_0/r_0$, which leads to $\sigma_R^2 = 1.637\ t^{5/6} (w_0/r_0)^{5/3}$. For $w_0/r_0=1$ we then find that $\sigma_R^2$ is proportional to $t^{5/6}$.

Based on the above argument it seems that one never reaches the strong scintillation region with a nonzero concurrence, regardless of the approximations used, which implies that the single phase screen approximation can be used for all situations. We must, however, emphasize that this conclusion is based on our current results, which only consider the case where only one photon propagates through turbulence and where the basis is restricted to $|\ell|= 1$. When both photons propagate through turbulence or when $|\ell|\gg 1$ the difference between the S\&R and IPE curves may become bigger, which may cause the scales for IPE and S\&R to deviate significantly. For such a case situations may exist where entanglement survives deeper into the strong scintillation region.

In summary, we present the first experimental confirmation of the theoretical decay of entanglement in atmospheric turbulence, where only one photon propagates through turbulence and the qubit basis is restricted to an OAM of $|\ell|=1$. More work remains to be done: the behavior for quantum states with larger OAM and the effect of different scenarios, such as when both photons propagate through turbulence, need further investigation.

This work was done with funding support from the CSIR.


\end{document}